\documentclass[conference]{IEEEtran}
\usepackage{amsmath,cite,amsfonts,amssymb,psfrag,amsthm,graphicx,paralist}
\usepackage{pgfplots}
\pgfplotsset{compat=1.7}
\usetikzlibrary{calc}
\usetikzlibrary{shapes,arrows}
\usetikzlibrary{decorations.markings}
\usepackage[justification=centering]{caption}
\usepackage{afterpage}
\usepackage{multirow}
\usepackage{url}
\usepackage{epstopdf}
\usepackage{gensymb}
\usepackage{nameref}

\usepackage{enumitem,afterpage}
\usepackage{dsfont}

\usepackage{setspace}
\newcommand{\Enc}{\mathsf{Enc}}
\newcommand{\Dec}{\mathsf{Dec}}

\newcommand*\xor{\mathbin{\oplus}}

\usepackage{array}
\makeatletter
\newcommand*\bigcdot{\mathpalette\bigcdot@{.5}}
\newcommand*\bigcdot@[2]{\mathbin{\vcenter{\hbox{\scalebox{#2}{$\m@th#1\bullet$}}}}}
\makeatother
\newcolumntype{M}[1]{>{\centering\arraybackslash}m{#1}}
\newcolumntype{N}{@{}m{0pt}@{}}

\makeatletter
\let\save@mathaccent\mathaccent
\newcommand*\if@single[3]{%
	\setbox0\hbox{${\mathaccent"0362{#1}}^H$}%
	\setbox2\hbox{${\mathaccent"0362{\kern0pt#1}}^H$}%
	\ifdim\ht0=\ht2 #3\else #2\fi
}
\newcommand*\rel@kern[1]{\kern#1\dimexpr\macc@kerna}
\newcommand*\widebar[1]{\@ifnextchar^{{\wide@bar{#1}{0}}}{\wide@bar{#1}{1}}}
\newcommand*\wide@bar[2]{\if@single{#1}{\wide@bar@{#1}{#2}{1}}{\wide@bar@{#1}{#2}{2}}}
\newcommand*\wide@bar@[3]{%
	\begingroup
	\def\mathaccent##1##2{%
		\let\mathaccent\save@mathaccent
		\if#32 \let\macc@nucleus\first@char \fi
		\setbox\z@\hbox{$\macc@style{\macc@nucleus}_{}$}%
		\setbox\tw@\hbox{$\macc@style{\macc@nucleus}{}_{}$}%
		\dimen@\wd\tw@
		\advance\dimen@-\wd\z@
		\divide\dimen@ 3
		\@tempdima\wd\tw@
		\advance\@tempdima-\scriptspace
		\divide\@tempdima 10
		\advance\dimen@-\@tempdima
		\ifdim\dimen@>\z@ \dimen@0pt\fi
		\rel@kern{0.6}\kern-\dimen@
		\if#31
		\overline{\rel@kern{-0.6}\kern\dimen@\macc@nucleus\rel@kern{0.4}\kern\dimen@}%
		\advance\dimen@0.4\dimexpr\macc@kerna
		\let\final@kern#2%
		\ifdim\dimen@<\z@ \let\final@kern1\fi
		\if\final@kern1 \kern-\dimen@\fi
		\else
		\overline{\rel@kern{-0.6}\kern\dimen@#1}%
		\fi
	}%
	\macc@depth\@ne
	\let\math@bgroup\@empty \let\math@egroup\macc@set@skewchar
	\mathsurround\z@ \frozen@everymath{\mathgroup\macc@group\relax}%
	\macc@set@skewchar\relax
	\let\mathaccentV\macc@nested@a
	\if#31
	\macc@nested@a\relax111{#1}%
	\else
	\def\gobble@till@marker##1\endmarker{}%
	\futurelet\first@char\gobble@till@marker#1\endmarker
	\ifcat\noexpand\first@char A\else
	\def\first@char{}%
	\fi
	\macc@nested@a\relax111{\first@char}%
	\fi
	\endgroup
}
\makeatother

\makeatletter
\newtheorem*{rep@theorem}{\rep@title}
\newcommand{\newreptheorem}[2]{%
	\newenvironment{rep#1}[1]{%
		\def\rep@title{\Cref{##1}}%
		\begin{rep@theorem}}%
		{\end{rep@theorem}}}
\newcommand*{\textlabel}[2]{%
	\edef\@currentlabel{#1}
	\phantomsection
	#1\label{#2}
}
\makeatother

\newtheorem{theorem}{Theorem}

\newtheorem{definition}{Definition}

\tikzset{XOR/.style={draw,circle,append after command={
			[shorten >=\pgflinewidth, shorten <=\pgflinewidth,]
			(\tikzlastnode.north) edge (\tikzlastnode.south)
			(\tikzlastnode.east) edge (\tikzlastnode.west)
		}
	}
}
\tikzset{line/.style={draw, -latex',shorten <=1bp,shorten >=1bp}}

\IEEEoverridecommandlockouts

\begin{document}
	\title{Quality of Service Guarantees for\\ Physical Unclonable Functions}
	\author{\IEEEauthorblockN{Onur G\"unl\"u\IEEEauthorrefmark{1}, Rafael F. Schaefer\IEEEauthorrefmark{1}, and H. Vincent Poor\IEEEauthorrefmark{2}}
		\IEEEauthorblockA{\IEEEauthorrefmark{1}Chair of Communications Engineering and Security, University of Siegen, 57076 Siegen, Germany \\
			Email: \{onur.guenlue, rafael.schaefer\}@uni-siegen.de}
		\IEEEauthorblockA{\IEEEauthorrefmark{2}Electrical and Computer Engineering Department, Princeton University, Princeton, NJ 08544, USA\\
			Email: poor@princeton.edu}
\thanks{O. G\"unl\"u and R. F. Schaefer were supported by the German Federal Ministry of Education and Research (BMBF) within the national initiative for ``Post Shannon Communication (NewCom)'' under the Grant 16KIS1004. The work of H. V. Poor was supported by the U.S. National Science Foundation (NSF) under Grants CCF-0939370, CCF-1513915, and CCF-1908308. }
}
	\maketitle
	
	\begin{abstract}
		We consider a secret key agreement problem in which noisy physical unclonable function (PUF) outputs facilitate reliable, secure, and private key agreement with the help of  public, noiseless, and authenticated storage. PUF outputs are highly correlated, so transform coding methods have been combined with scalar quantizers to extract uncorrelated bit sequences with reliability guarantees. For PUF circuits with continuous-valued outputs, the models for transformed outputs are made more realistic by replacing the fitted distributions with corresponding truncated ones. The state-of-the-art PUF methods that provide reliability guarantees to each extracted bit are shown to be inadequate to guarantee the same reliability level for all PUF outputs. Thus, a quality of service parameter is introduced to control the percentage of PUF outputs for which a target reliability level can be guaranteed. A public ring oscillator (RO) output dataset is used to illustrate that a truncated Gaussian distribution can be fitted to transformed RO outputs that are inputs to uniform scalar quantizers such that reliability guarantees can be provided for each bit extracted from any PUF device under additive Gaussian noise components  by eliminating a small subset of PUF outputs. Furthermore, we conversely show that it is not possible to provide such reliability guarantees without eliminating any PUF output if no extra secrecy and privacy leakage is allowed.
	\end{abstract}
	\IEEEpeerreviewmaketitle

\section{Introduction}
Authentication and identification of devices in a digital network are necessary to protect sensitive data. One classic method assigns a unique identifier to each device, in which an identifier can either be inserted by a trusted party or extracted from each device separately. Similar to biometric identifiers that identify an individual, physical identifiers such as physical unclonable functions (PUFs) \cite{GassendThesis} are used to uniquely and reliably identify a device that embodies the physical identifier. Applications of PUFs include securing internet-of-things (IoT) devices that carry personal data by using PUFs embodied in these devices to store secret keys (SKs) that can be reconstructed on demand by using the PUF outputs \cite{benimdissertation}. Thus, high-entropy and reliable circuit outputs in a device, such as oscillation frequencies of ring oscillators (ROs), are used as PUFs that are safer and cheaper alternatives to storing SKs in a non-volatile memory \cite{ROFirst,benimthesis}. PUFs are safer since the SK is reconstructed only on demand and an invasive attack changes the PUF outputs permanently, both of which eliminate the need for costly continuous hardware protection \cite{pufintheory}.

Measurements of digital circuit outputs are \emph{noisy} mainly due to random temporal variations in the hardware, and there are \emph{correlations} between different digital circuit outputs that are embodied in the same device mainly due to surrounding logic circuitry that causes systematic variations; see \cite{ROFirst,MerliROCorrelated} for such discussions about RO PUFs. The noise in  PUF output measurements result in errors in the SK extracted, which should be corrected by using error correcting codes (ECCs) \cite{ROFirst}. Furthermore, correlations between different PUF output symbols might cause extra information leakage about the SK to an eavesdropper that knows the correlation since the eavesdropper can apply, e.g., machine learning algorithms to model the PUF outputs \cite{MLPUF,bizimMDPI}. For PUF measurement channels with additive noise components, SK agreement schemes that use ECCs, also called \textit{helper data schemes}, are proposed. The fuzzy commitment scheme (FCS)\cite{FuzzyCommitment} and the code-offset fuzzy extractors \cite{Dodis2008fuzzy} are classic helper data schemes; see \cite{ourISITATrifonov,bizimWZ} for their extensions with a limit on the amount of storage. However, these helper data schemes require the PUF outputs to be independent and identically distributed (i.i.d.) according to a uniform probability distribution to achieve the SK capacity, which is equal to the maximum achievable ECC rate \cite{IgnatenkoFuzzy, Maurer,AhlswedeCsiz,IgnaTrans}. Thus, in \cite{bizimMDPI,bizimpaper}, and \cite{OurICASSP2020} transform coding methods are applied to correlated PUF circuit outputs to obtain highly-uncorrelated transform coefficients that can be quantized separately by using uniform scalar quantizers in order to extract almost i.i.d. and uniformly distributed outputs in a transform domain, as being applied to biometric identifiers~\cite{Campisi,Transformbio,BorisTransform}; see \cite{MLbasedTransform,VQTransform} for more complex alternative methods. 

Suppose the noisy and correlated PUF circuit outputs are continuous-valued, as they are for RO PUFs. We extract SKs from such PUFs by applying a new transform coding method that significantly improves the state-of-the-art methods by
  \begin{inparaenum}
	\item fitting truncated probability distributions to the noiseless transform coefficients obtained from noiseless PUF circuit outputs, such as RO outputs, to take into consideration that most of digital circuits put out values within a finite range;
	\item introducing a quality of service (QoS) parameter that determines the percentage of PUF outputs for which a target reliability level can be guaranteed;
	\item determining the effects of the QoS parameter on the trade-off between the number of bits one can extract from a transform coefficient and the average or maximum error probability with Gray labeling, Gaussian distributed transform coefficients, and additive Gaussian noise components;
	\item proving the existence of better schemes than the classic helper data schemes by illustrating that the measurement channel model after a uniform scalar quantization is generally not memoryless;
	\item evaluating the results for RO PUFs to illustrate the QoS gains in terms of reliability at a cost of removal of a small subset of PUF outputs, in which we apply low-complexity state-of-the-art orthogonal transforms that do not require any multiplications.
\end{inparaenum}

\section{RO Output Model}\label{sec:ROModel}
We briefly describe the digital circuit model for ROs since the RO PUF is a classic PUF type that uses continuous- and positive-valued circuit outputs for SK agreement and we focus on these PUFs in this work. The logic circuit of an RO consists of serially-connected odd number of inverters in which the output of the last inverter is fed back as the input of the first inverter. We remark that the first logic gate is generally chosen to be a NAND gate that provides the same logic output as an inverter if the NAND gate is enabled and that allows to disable the RO when unused. The manufacturing-dependent and uncontrollable component in an RO is the total propagation delay of an input signal to flow through the RO, which determines the oscillation frequency of an RO. This component is used as the source of randomness for RO PUFs. 

Measurements of an RO output are noisy due to random noise sources, such as flicker noise and thermal noise, and different RO outputs are correlated due to deterministic effects, such as the cross-talk between signal traces that are adjacent and surrounding logic circuitry in the hardware \cite{RingOscillators,pufintheory,benimthesis}. The traditional method \cite{ROFirst} to extract binary SKs from RO outputs makes hard binary decisions by comparing the oscillation frequencies of RO pairs. However, correlations between outputs of different ROs, as discussed above, cause extra secrecy leakage \cite{regressdistiller}. Therefore, a discrete cosine transform (DCT) based transform coding scheme is proposed in \cite{bizimpaper} to reduce the correlations before applying scalar quantizers. The discrete Walsh Hadamard transform (DWHT) is shown in \cite{bizimMDPI} to both achieve a similar decorrelation performance as being achieved by the DCT for a public RO output dataset \cite{ROPUF} and  require a significantly smaller hardware area than being required for the DCT. Similarly, a new set of orthogonal transforms, including the DWHT, is proposed in \cite{OurICASSP2020} to illustrate the possibility to reduce the error probability in the transform domain without increasing the hardware complexity compared to the DWHT. Thus, in our transform coding method we use the new set of orthogonal transforms \cite{OurICASSP2020} and publicly select the transform in the set that achieves the best decorrelation and reliability performance with QoS guarantees.

\section{A Secret Key Agreement Method}\label{sec:fuzzycommitment}
 Suppose a SK is bound to a digital circuit output. The digital circuit outputs can be used as a PUF if the SK can be reconstructed reliably by measuring the same circuit outputs again, which is possible for a large set of digital circuit output and noise distributions by using an ECC \cite{Maurer,AhlswedeCsiz}. We describe the FCS that uses an ECC and a masking step to reconstruct the same SK reliably by using a noisy measurement of the same digital circuit output. Without loss of generality, we assume that the first PUF output measurement is noiseless and other measurements are noisy; see \cite[Appendix~B]{bizimWZ} for the steps to extend these results to the case where the PUF output is hidden (or remote) such that all of its measurements are noisy.

Denote the first measurement of a PUF output as an $n$-letter sequence $X^n\in\mathcal{X}^n$. A pre-determined SK $S\in\mathcal{S}$ is embedded to bind $S$ to $X^n$ such that the second PUF output measurement $Y^n\in\mathcal{Y}^n$ and the output of the binding operation suffice to reconstruct $S$ reliably. The FCS satisfies the reliability constraint by publicly storing $n$-letter \emph{helper data} $W^n\in\mathcal{W}^n$ that are obtained by applying a masking step that takes two $n$-letter sequences as its inputs. Consider a linear ECC $\mathbb{C}$ with blocklength $n$, code dimension $\log|\mathcal{S}|$, encoder $\Enc(\cdot)$, and decoder $\Dec(\cdot)$. The masking step calculates modulo-$|\mathcal{X}|$ sum of $X^n$ and a codeword $C^n$ that is obtained from the SK $S$ by encoding it as $C^n=\Enc(S)$. The most common FCS assumes that $\mathcal{X}=\mathcal{Y}=\mathcal{W}=\{0,1\}$ such that a binary linear ECC can be used, which is assumed also below for simplicity. Thus, we have the helper data $W^n=X^n\xor C^n$, where $\xor$ represents modulo-2 sum. Similarly, one can represent the second PUF measurement sequence $Y^n$ as a noisy version of the first measurement $X^n$ such that $Y^n=X^n\xor E^n$, where $E^n\in\{0,1\}^n$ represents a binary error sequence. We have $W^n\xor Y^n = C^n\xor E^n$, so the decoder $\Dec(\cdot)$ can decode the noisy codeword sequence $C^n\xor E^n$ into an index $\hat{S}\in\mathcal{S}$. The FCS assumes that $X^n$ is distributed according to an i.i.d. Bernoulli distribution $P_X$ and the channel $P_{Y^n|X^n}$ is memoryless, i.e., $P_{Y^n|X^n}=P^n_{Y|X}$. These assumptions are necessary (but not sufficient) to achieve a rate tuple that is on the rate region boundary by using the FCS \cite{FuzzyCommitment}. We next define the rate region that consists of all achievable (secret-key, privacy-leakage) rate pairs by using the FCS under reliability, secrecy, and privacy constraints. Moreover, we illustrate a binary input distribution $P_X$ and a memoryless channel $P_{Y|X}$ for which the FCS is asymptotically optimal.

\begin{definition}
	\normalfont A (secret-key, privacy-leakage) rate pair $ (R_\text{s}\text{,}R_\ell)$ is \emph{achievable} by using the FCS if, for any $\epsilon\!>\!0$, there exist $n\!\geq\!1$, an encoder $\Enc(\cdot)$, and a decoder $\Dec(\cdot)$ that satisfy $R_\text{s}=\log|\mathcal{S}|/n$ and
	\begin{alignat}{2}
	&\Pr[\hat{S}\ne S] \leq \epsilon && (\text{reliability}) \label{eq:reliabilityconst}\\
		&H(S)\geq n(R_{\text{s}}-\epsilon)&&(\text{SK uniformity})\label{eq:secretkeyuniformity}\\
	&I\big(W^n;S\big)=0 && (\text{zero secrecy leakage})\label{eq:secrecyconst}\\
	&I\big(W^n;X^n\big) \leq n(R_\ell+\epsilon) \quad\quad\quad&&(\text{privacy-leakage rate})  \label{eq:privacyconst}.
	\end{alignat}
\end{definition}

An achievable rate pair $(R_\text{s}\text{,}R_\ell)$ should thus satisfy the conditions that the probability of error in reconstructing the SK $S$ is negligible (\ref{eq:reliabilityconst}), the SK $S$ with rate $R_{\text{s}}$ should be almost uniformly distributed (\ref{eq:secretkeyuniformity}), the only public sequence $W^n$ that is available to an eavesdropper should not leak any information about $S$ to achieve \emph{perfect secrecy} (\ref{eq:secrecyconst}), and the normalized amount of information leaked to an eavesdropper about the first PUF output measurement $X^n$ should not be non-negligibly larger than the privacy-leakage rate $R_\ell$, which is motivated by multiple PUF enrollments. We remark that the unnormalized privacy-leakage metric $I(W^n;X^n)$ is in general unbounded unless, e.g., a private key is available during all PUF output measurements \cite{IgnaTrans}, which is not practical because if a private key that is hidden from an eavesdropper is available, then there is no need to bind SKs to PUF output measurements; see \cite{bizimITW} for scenarios in which strong privacy can be achieved without a private key.  

A classic measurement channel $P_{Y|X}$ model for PUFs is a binary symmetric channel (BSC) with crossover probability $0\leq p\leq 1$, which is used, e.g., for static random access memory (SRAM) PUFs \cite{SRAMPUFFirst}. In the next section, we describe the new transform coding method and illustrate that a BSC fits well to the measurement channel $P_{Y|X}$ model also for PUF circuits with continuous-valued outputs, such as RO PUFs, if uniform scalar quantizers are applied after transformation and all noise components as well as transform coefficients have symmetric probability distributions. 

The analysis of the FCS assumes that $X^n$ is i.i.d. Thus, one constraint in choosing the orthogonal transform that is applied to PUF circuit outputs is that its decorrelation performance should be good such that almost i.i.d. PUF output symbols can be extracted in the transform domain by using scalar quantizers. Similarly, the channel $P_{Y^n|X^n}$ is assumed to be memoryless for the FCS analysis. We show below that these assumptions can be satisfied for RO PUFs by applying the new transform coding method since we obtain PUF output measurements $X^n$ that are almost i.i.d. according to a binary uniform distribution and have $P_{Y^n|X^n}\simeq\prod_{i=1}^{n}P_{Y_i|X_i}$, where $P_{Y_i|X_i}$ a BSC with crossover probability $p$ for all $i\in [1:n]$. We next illustrate the region of all achievable (secret-key, privacy-leakage) rate pairs for this case. Define the binary entropy function $\displaystyle H_b(p)\!=\!-p\log p-(1\!-p)\log(1\!-p)$.

\begin{theorem}[\hspace{1sp}\cite{IgnatenkoFuzzy}]\label{theo:BSSBSCFCS}
	The region of all achievable rate pairs $(R_\text{s}\text{,}R_\ell)$ for the FCS with i.i.d. $X^n$, binary uniform distribution $P_X$, and memoryless measurement channel $P_{Y|X}$ that is a BSC with crossover probability $p\in[0,1]$ is 
	\begin{align}
	\big\{ \left(R_\text{s},R_\ell\right)\!\colon\!\quad 0\leq R_\text{s}\leq 1-H_b(p),\quad R_\ell\geq 1\!-\!R_\text{s} \big\}.\label{eq:ls0}
	\end{align}
\end{theorem}

The FCS is asymptotically optimal only at the rate tuple $(R^*_{\text{s}},R^*_\ell) = (1\!-\!H_b(p), H_b(p))$ \cite{IgnatenkoFuzzy,IgnaTrans}. Since the maximum achievable SK rate $R^*_{\text{s}}$ is equal to the channel capacity of the channel $P_{Y|X}$, it suffices to maximize the rate of the ECC $\mathbb{C}$ to achieve asymptotic optimality with the FCS. Thus, we next focus only on providing QoS guarantees by proposing a new transform coding method to obtain output models that follow the probability distributions given in Theorem~\ref{theo:BSSBSCFCS} since the ECC design for SK agreement with the FCS can be handled by using techniques proposed in \cite{bizimpaper,bizimICC}.

\section{Proposed Transform Coding Steps}\label{sec:commonsteps}
We describe our new transform coding method that can be applied to a large set of PUF circuits with continuous-valued outputs and, for simplicity, we focus on RO PUFs to analyze the performance of the proposed transform coding method. Suppose $r$ ROs, where $\sqrt{r}\!\in\! \mathbb{Z}^+$, are implemented as a two-dimensional (2D) array of size $\sqrt{r}\!\times\! \sqrt{r}$ and the first RO output measurements are represented as a vector random variable $\widetilde{X}^{r}$ that is distributed according to a joint probability density function $\displaystyle f_{\widetilde{X}^{r}}$, i.e., we allow correlations between symbols of $\widetilde{X}^{r}$. Suppose additive random noise sequence $\widetilde{E}^r$ consists of symbols with zero mean and denote the second RO output measurement as $\widetilde{Y}_j\!=\!\widetilde{X}_j \!+\!\widetilde{E}_j$ for all $j\!\in\![1\!:\!r]$. We next describe the new transform coding method that applies a 2D $\sqrt{r}\!\times\! \sqrt{r}$ orthogonal transform to the RO output measurements and then scalar quantizers to extract binary sequences $X^n$ and $Y^n$ from the first and second RO output measurements, respectively. Denote the binary error symbols in the transform domain as $E_i=X_i\xor Y_i$ for all $i\in[1:n]$. 

The new transform coding method consists of the following steps:
\begin{inparaenum} 
\item a 2D transformation to decorrelate $r$ RO output measurements;
\item modeling noiseless transform coefficients and additive noise components for realistic analysis;
\item histogram equalization to convert all noiseless transform coefficients into realizations of random variables with fixed mean and variance values such that low-complexity scalar quantizers with simpler analysis can be applied;
\item scalar uniform quantization of each transform coefficient to obtain an almost i.i.d. and uniformly distributed binary sequence $X^n$ by applying Gray labeling and concatenating all bits extracted from used transform coefficients. We remark that in the last step we impose the QoS constraint for the new equalized transform coefficient probability distribution model.   
\end{inparaenum}
\vspace*{-0.0cm}
\subsubsection{Transformation}\label{subsec:Step 1}
The main aim of the transformation step is to decorrelate RO output measurements $\widetilde{X}^r$ such that transform coefficients $\widetilde{T}^r$ that are obtained from $\widetilde{X}^r$ can be quantized separately with a negligible loss in security as the transform coefficients are \emph{mutually independent} if they are uncorrelated and jointly Gaussian distributed. Using transforms that put out almost independent transform coefficients is common in, e.g., the digital watermarking and image processing literature \cite{mySPbook}. In a $\sqrt{r}\!\times\!\sqrt{r}$ RO output array, the neighboring outputs are observed to be highly-correlated \cite{MerliROCorrelated}. Moreover, the decorrelation performance of a transform can be measured by the \emph{decorrelation efficiency} metric \cite{decorrelationbook}, which is determined by the ratio of the sums of absolute values of non-diagonal elements in the autocovariance matrices that are calculated before and after transformation. The maximum decorrelation efficiency is achieved by using the Karhunen-Lo\`eve transform (KLT) for a large set of probability distributions, but the KLT has high computational complexity. Low-complexity 2D $\sqrt{r}\!\times\!\sqrt{r}$ transforms with high decorrelation efficiency are proposed and implemented in \cite{bizimMDPI} for SK agreement with RO PUFs that are used in IoT applications. Proposed low-complexity transforms include the DWHT and its extensions are obtained in \cite[Section~4.1]{OurICASSP2020} by exhaustively searching all orthogonal matrices of size $4\!\times \!4$ with matrix elements from the set $\{-1,1\}$ and then constructing larger matrices by using one of the orthogonal matrices multiple times such that orthogonality is preserved for the larger matrix. The new set of transforms consists of $12288$ orthogonal transforms for a transform size $16\!\times\! 16$, each of which can be implemented without multiplications with a negligible performance loss in terms of the decorrelation efficiency as compared to the DCT \cite{bizimMDPI,OurICASSP2020}. The transform, among these orthogonal transforms, that achieves the minimum value for the error probability maximized over all used transform coefficients obtained from the RO output dataset \cite{ROPUF} is used for analysis in \cite{OurICASSP2020}, and it is called the \emph{selected transform (ST)}. Thus, in Section~\ref{sec:effectsofQoSforROPUFs} we also use the ST for our RO PUF reliability analysis.
\vspace*{-0.05cm}
\subsubsection{Modeling Transform Coefficients and Noise}\label{subsec:Step2}
Consider the transform coefficients $\widetilde{T}^r$ that are obtained from the RO output measurements $\widetilde{X}^r$ in the dataset \cite{ROPUF} by applying a transform in the set of orthogonal transforms proposed in \cite[Section~4.1]{OurICASSP2020}. Distribution fitting criteria used in  \cite{OurICASSP2020} suggest that a Gaussian distribution can be fitted to all transform coefficients. However, each RO output realization $\widetilde{x}_j$ takes on a value from a finite range that has to consist of positive real numbers since RO outputs are oscillation frequencies and that depends on the  technology node used to implement the RO logic circuit. Therefore, we fit a truncated Gaussian distribution to each used transform coefficient $\widetilde{T}_j$, i.e., for all $j\!\in\![2\!:\!r]$ since the DC coefficient $\widetilde{T}_1$ corresponds to the average oscillation frequency over $r$ ROs and its value can be estimated reliably by an attacker \cite{benimthesis}. Unbiased mean and variance parameters of the fitted distributions are estimated via maximum-likelihood estimation and the finite ranges are determined from the transform coefficients obtained from the RO output dataset \cite{ROPUF}. Furthermore, we apply the same 2D $\sqrt{r}\times\sqrt{r}$ transform that is applied to $\widetilde{X}^r$ also to the second RO output measurements $\widetilde{Y}^r$. The transform coefficients obtained from $\widetilde{Y}^r$ can be represented as \emph{noisy transform coefficients} $(\widetilde{T}_j+\widetilde{N}_j)$ for all $j\in[2:r]$ such that the additive noise components $\widetilde{N}_j$ are mutually independent and zero-mean Gaussian distributed as well as independent of $\widetilde{T}^r$. Noise variances estimated from the RO output dataset \cite{ROPUF} are small compared to the finite range of the truncated Gaussian distributions. Thus, a truncated version of the Gaussian distributed noise components result in negligible differences in the reliability analysis, so we use the Gaussian distribution as the distribution of $\widetilde{N}_j$ for all $j\in[2\!:\!r]$ for simplicity.
\vspace*{-0.05cm}
\subsubsection{Histogram Equalization}
The histogram equalization step is proposed first in \cite{bizimGlobalSIP} to convert each transform coefficient $\widetilde{T}_j$ that is modeled in \cite{bizimGlobalSIP} as a Gaussian distribution with mean $\mu_{\widetilde{T}_j}\neq 0$ and variance $\sigma^2_{\widetilde{T}_j}\neq 1$ into a standard Gaussian distribution for all $j\in[2:r]$. This step simplifies the error probability analysis, so we apply a similar histogram equalization step. First, consider the original Gaussian distribution with mean $\mu_{\widetilde{T}_{j,\text{orig}}}$ and variance $\sigma^2_{\widetilde{T}_{j,\text{orig}}}$ from which the truncated Gaussian distribution that is fitted to the $j$-th transform coefficient $\widetilde{T}_j$ is obtained by bounding its range from both above and below. The mean and variance of the truncated Gaussian distribution is uniquely determined by $\mu_{\widetilde{T}_{j,\text{orig}}}$, $\sigma^2_{\widetilde{T}_{j,\text{orig}}}$, and the lower and upper bounds on its range \cite{JohnsonTruncatedGauss}. Thus, as the modified histogram equalization step that simplifies the analysis, from each realization $\widetilde{T}_j=\widetilde{t}_j$ we first subtract $\mu_{\widetilde{T}_{j,\text{orig}}}$ and then divide the result by $\sigma_{\widetilde{T}_{j,\text{orig}}}$ by enforcing the equalized original Gaussian distribution to be a standard Gaussian distribution. Denote the resulting equalized transform coefficient, distributed according to a truncated Gaussian distribution, as $\widebar{\widetilde{T}}_j$ and the resulting additive zero-mean mutually-independent Gaussian noise component with variance $\sigma^2_{\widebar{\widetilde{N}}_j}$ as $\widebar{\widetilde{N}}_j$, respectively, for all $j\in[2:r]$. 
\vspace*{-0.03cm}
\subsubsection{Quantizing Noisy Transform Coefficients for Reliable Bit Extraction with QoS}\label{subsec:Step4}	
Suppose we extract $m_j\geq 0$ mutually independent and uniformly distributed bits from an equalized transform coefficient $\widebar{\widetilde{T}}_j$ for $j\in[2:r]$ such that the FCS can be used with almost i.i.d. and uniformly distributed binary sequences $X^n$. A sequence $x^n$ is obtained by concatenating the bit sequences extracted from $(r-1)$ equalized transform coefficients, so we have $n\!=\!\sum_{j=2}^{r}m_j$. Denote quantization boundaries of the $j$-th uniform scalar quantizer as $b_{j,0},b_{j,1},\ldots, b_{j,2^{m_j}}$, where $b_{j,0}$ and $b_{j,2^{m_j}}$ are lower and upper bounds on the range of $\widebar{\widetilde{T}}_j$, respectively. For all $j\in[2\!:\!r]$ and $k_j\in[1\!:\!(2^{m_j}\!-\!1)]$ we assign the quantiles of the $j$-th equalized and truncated Gaussian distribution to the quantization boundaries, i.e., we have
\begin{align}
	b_{j,k_j} \!=\! Q^{-1}\left(Q(b_{j,0})\!\cdot\!\left(1\!-\!\frac{k_j}{2^{m_j}}\right)+Q(b_{j,2^{m_j}})\!\cdot\!\frac{k_j}{2^{m_j}}\right)\label{eq:quantizationboundaries}
\end{align}	
where $Q(\cdot)$ is the $Q$-function. Given any realization $\widebar{\widetilde{t}}_j$, or its noisy version $(\widebar{\widetilde{t}}_j\!+\!\widebar{\widetilde{n}}_j)$, this quantizer outputs $k_j$ if $\displaystyle b_{j,(k_j-1)}\!<\!\widebar{\widetilde{t}}_j\!\leq\!b_{j,k_j}$. Moreover, since each additive noise component $\widebar{\widetilde{N}}_j$ has zero mean, Gray labeling is applied to map each $k_j$ to a bit sequence of size $m_j$ for all $j\in[2:r]$ because this labeling results in only one bit flip if a noisy transform coefficient is quantized into a neighboring quantization interval.

\paragraph{QoS Analysis}
Suppose the observed realization of a transform coefficient is equal to a quantization boundary, i.e., $\widebar{\widetilde{t}}_j=b_{j,k_j}$ for some $j\in[2:r]$ and $k_j\in[1:(2^{m_j}\!-\!1)]$. Because of zero-mean independent additive Gaussian noise, the error probability for such a realization with $1$-bit quantization is $0.5$, so reliable reconstruction of the bit sequence mapped to the quantizer output $k_j$ is not possible; see \cite{SlavMostRelevant,SlavSecondMostRelevant,SlavPolarization} for similar discussions with different design metrics and without QoS guarantees. We remark that every set of equalized transform coefficient realizations $(\widebar{\widetilde{t}}_2,\widebar{\widetilde{t}}_3,\ldots,\widebar{\widetilde{t}}_r)$ corresponds to a 2D array of $r$ ROs that are embodied in a digital device and that are used as a PUF. Thus, in order to provide reliability guarantees to each RO PUF output, it is necessary to eliminate such unreliable realizations before quantization, which are the ones spatially close to the quantization boundaries. We provide such guarantees by eliminating the realizations $\widebar{\widetilde{t}}_j\in ((b_{j,k_j}\!-\!\delta/2),\, (b_{j,k_j} \!+\!\delta/2)]$ for all $j\in[2:r]$ and $k_j\in[1:(2^{m_j}\!-\!1)]$, and for some fixed $\delta\geq 0$, so the parameter $\delta$ is a \emph{QoS parameter} for all PUF outputs that are used for SK agreement with the FCS. Denote the ratio of eliminated realizations vs. all realizations for all $j\in[2:r]$ as
\begin{align}
	\gamma_j(\delta) = \frac{\displaystyle\sum_{k_j=1}^{(2^{m_j}\!-\!1)}\Big(Q\Big(b_{j,k_j}\!-\!\frac{\delta}{2}\Big)\!-\!Q\Big(b_{j,k_j}\!+\!\frac{\delta}{2}\Big)\Big)}{Q(b_{j,0})-Q(b_{j,2^{m_j}})}.\label{eq:gammaj}
\end{align}
For a fixed $\delta$, the percentage $\beta_j$ of realizations $\widebar{\widetilde{t}}_j$ that can be used for SK agreement is $\beta_j(\delta)\!=\!100\!\times\!(1\!-\!\gamma_j(\delta))$ for all $j\!\in\![2\!:\!r]$, decreasing for increasing $\delta$. The \emph{worst case error probability} decreases from $0.5$ to $ Q(\delta/2\sigma_{\widebar{\widetilde{N}}_j})$ for $1$-bit quantization, so $\delta$ represents a worst case reliability guarantee. 

We next illustrate that the error probabilities of different bits extracted from the same coefficient are dependent, i.e., the channel $P_{Y|X}$ is not memoryless. This proves that the FCS that requires $P_{Y|X}$  
to be memoryless, as discussed in Section~\ref{sec:fuzzycommitment}, can be improved by taking into consideration the memory in the channel. Assume, e.g., $m_j=2$ bits are extracted from $\widebar{\widetilde{T}}_j$ by applying a binary-reflected Gray labeling, i.e., the quantization intervals are mapped to ``00", ``01", ``11", ``10" in the given order. Then, we have 
\begin{align*}
	&\Pr\!\left[\!\{\text{1st bit is in error}\}\Big|\widebar{\widetilde{t}}_j\!\right]\!\!\cdot\!\left(Q(b_{j,0})\!-\!Q(b_{j,2^{m_j}})\right)\!\cdot\!(1\!-\!\gamma_j(\delta))\nonumber\\
	&=\begin{cases}
			Q\Big(\frac{b_{j,2}-\widebar{\widetilde{t}}_j}{\sigma_{\widebar{\widetilde{N}}_j}}\Big) & \text{if    }\; \widebar{\widetilde{t}}_j\in[b_{j,0},\,(b_{j,2}\!-\!\frac{\delta}{2})]\\
			Q\Big(\frac{\widebar{\widetilde{t}}_j-b_{j,2}}{\sigma_{\widebar{\widetilde{N}}_j}}\Big) & \text{if }\; \widebar{\widetilde{t}}_j\in((b_{j,2}\!+\!\frac{\delta}{2}),\,b_{j,4}],
	\end{cases}
\end{align*}
\vspace*{-0.2cm}
\begin{align*}
&\Pr\!\left[\!\{\text{2nd bit is in error}\}\Big|\widebar{\widetilde{t}}_j\!\right]\!\!\cdot\!\!\left(Q(b_{j,0})\!-\!Q(b_{j,2^{m_j}}\!)\right)\!\!\cdot\!(1\!-\!\gamma_j(\delta))\nonumber\\
&=\begin{cases}
\!Q\Big(\frac{b_{j,1}-\widebar{\widetilde{t}}_j}{\sigma_{\widebar{\widetilde{N}}_j}}\Big) \!\!-\!Q\Big(\frac{b_{j,3}-\widebar{\widetilde{t}}_j}{\sigma_{\widebar{\widetilde{N}}_j}}\Big)&\!\!\!\text{if } \widebar{\widetilde{t}}_j\!\in\![b_{j,0},\,(b_{j,1}\!-\!\frac{\delta}{2})]\\
\!Q\Big(\frac{\widebar{\widetilde{t}}_j-b_{j,1}}{\sigma_{\widebar{\widetilde{N}}_j}}\Big) \!\!+\! Q\Big(\frac{b_{j,3}-\widebar{\widetilde{t}}_j}{\sigma_{\widebar{\widetilde{N}}_j}}\Big)&\!\!\!\text{if } \widebar{\widetilde{t}}_j\!\in\!((b_{j,1}\!+\!\frac{\delta}{2}),\,(b_{j,3}\!-\!\frac{\delta}{2})]\\
\!Q\Big(\frac{b_{j,1}-\widebar{\widetilde{t}}_j}{\sigma_{\widebar{\widetilde{N}}_j}}\Big) \!\!-\! Q\Big(\frac{b_{j,3}-\widebar{\widetilde{t}}_j}{\sigma_{\widebar{\widetilde{N}}_j}}\Big)&\!\!\!\text{if } \widebar{\widetilde{t}}_j\!\in\!((b_{j,3}\!+\!\frac{\delta}{2}),\;b_{j,4}].
\end{cases}
\end{align*}
Applying the Bayes' theorem and the law of total probability to $\Pr[\{\text{1st bit is in error}\}|\widebar{\widetilde{t}}_j]$ (or $\Pr[\{\text{2nd bit is in error}\}|\widebar{\widetilde{t}}_j]\,$) given above, one can obtain for each quantization interval separately the formula for the probability of the first (or second) bit being erroneous conditioned on the event that the equalized transform coefficient $\widebar{\widetilde{t}}_j$ falls into the corresponding quantization interval. Since closed form expressions do not seem to exist for these probabilities, we compute them numerically for each quantization interval for various parameters and observe that the multiplication of these two marginal probabilities is generally not equal to the corresponding joint probability. Thus, we numerically prove that the errors in the first and second bits conditioned on a quantization interval, which determines the mapped bit sequence, are dependent, i.e., the channel $P_{Y|X}$ is in general not memoryless, so it is not optimal to use the FCS; see \cite{Anesthesis} for error probability analysis without truncation or QoS for $m_j\geq 2$. 

An alternative reliability metric $P_{c}$, called \emph{correctness probability}, that measures the probability of the event that all extracted bits are correct is proposed in \cite{bizimICC} as a conservative metric that can be used in combination with the FCS. Furthermore, ECC $\mathbb{C}$ design procedures are proposed in \cite{bizimICC,bizimMDPI} that apply a thresholding approach to $P_{c}$ of each transform coefficient, in which a bounded minimum distance decoder which can correct a fixed number of symbol errors is assumed to determine how many bits should be extracted from each coefficient to satisfy a given block error probability constraint. Slight modifications of these design procedures can be applied to our transform coding steps as well, so we focus only on the correctness probability calculations for our transform coding method with more realistic models and QoS guarantees. For an equalized transform coefficient $\widebar{\widetilde{T}}$ with QoS parameter $\delta$, we have the following result, where the index $j$ is omitted.
\begin{align}
 &P_{c}(\delta)\cdot\left(Q(b_{0})-Q(b_{2^{m}})\right)\cdot(1-\gamma(\delta))\nonumber\\
 &=\textstyle\int\displaylimits_{b_{0}}^{(b_{1}-\delta/2)}\Big[\!Q\Big(\frac{b_{0}\!-\!\widebar{\widetilde{t}}}{\sigma_{\widebar{\widetilde{N}}}}\Big)\!-\!Q\Big(\frac{b_{1}\!-\!\widebar{\widetilde{t}}}{\sigma_{\widebar{\widetilde{N}}}}\Big)\!\Big]f_{\widebar{T}}(\widebar{\widetilde{t}})d{\widebar{\widetilde{t}}}\nonumber\\
 &\;\; +\!\!\! \sum_{k=1}^{(2^{m}-2)}\!\!\textstyle\int\displaylimits_{(b_{k}+\delta/2)}^{(b_{(k+1)}-\delta/2)}\Big[\!Q\Big(\frac{b_{k}\!-\!\widebar{\widetilde{t}}}{\sigma_{\widebar{\widetilde{N}}}}\Big)\!-\!Q\Big(\frac{b_{(k+1)}\!-\!\widebar{\widetilde{t}}}{\sigma_{\widebar{\widetilde{N}}}}\Big)\!\Big]f_{\widebar{T}}(\widebar{\widetilde{t}})d{\widebar{\widetilde{t}}}\nonumber\\
 & \;\;+ \textstyle\int\displaylimits_{(b_{(2^{m}-1)}+\delta/2)}^{b_{2^m}}\!\Big[Q\Big(\frac{b_{(2^{m}-1)}\!-\!\widebar{\widetilde{t}}}{\sigma_{\widebar{\widetilde{N}}}}\Big)\!-\!Q\Big(\frac{b_{2^m}\!-\!\widebar{\widetilde{t}}}{\sigma_{\widebar{\widetilde{N}}}}\Big)\!\Big]f_{\widebar{T}}(\widebar{\widetilde{t}})d{\widebar{\widetilde{t}}}\label{eq:P_cformula}
\end{align}
where $\widebar{T}$ is a random variable that is distributed according to a standard Gaussian probability density function $f_{\widebar{T}}$.

\section{Effects of QoS for RO PUFs and Discussions}\label{sec:effectsofQoSforROPUFs}
\begin{figure}[t]
	\centering
	\vspace*{-2.32cm}
    \input{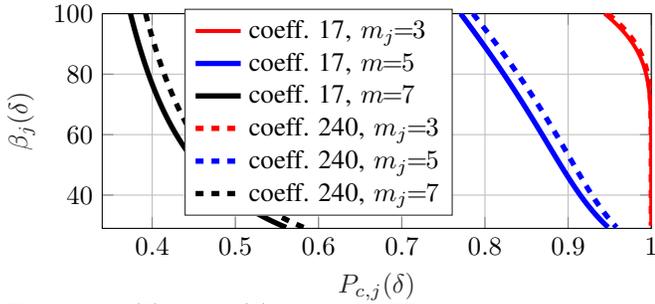}
	\vspace*{-0.85cm}
	\caption{$P_{c,j}(\delta)$ vs. $\beta_j(\delta)$ when the ST is applied to $16\times16$ RO arrays. We have $\beta_j=100$ when  $\delta=0$, and $\beta_j$ decreases for increasing $\delta$. Coefficient $j$ is the transform coefficient in row $\lceil j/16\rceil$ and column $(j\!\!\!\mod16)$.} 
	\label{fig:PcvsUsablePercentage}
	\vspace*{-0.5cm}
\end{figure}

We use the public RO output dataset \cite{ROPUF}, consisting of $100$ noisy measurements of $32\times16$ RO output arrays obtained from $193$ different devices, but we consider only the upper part of the array such that $\sqrt{r}=16$ to apply the transform coding steps described in Section~\ref{sec:commonsteps}. In Step 1, we apply the ST to the $16\times 16$ RO array. Applying Steps 2-4, we compute $\beta_j(\delta)$ by using (\ref{eq:gammaj}) and $P_{c,j}(\delta)$ from (\ref{eq:P_cformula}). We plot in Fig.~\ref{fig:PcvsUsablePercentage} the effects of $\delta$ on tuples $(P_{c,j},\;\beta_j)$ for two randomly-chosen transform coefficients that are uniformly quantized by using three different bit sequence lengths, i.e., $m_j\!=\!3,5,7$.

When $\delta$ increases, the percentage of realizations that can be used decreases, whereas the correctness probability increases as depicted in Fig.~\ref{fig:PcvsUsablePercentage}. The allowed range of values for $\delta$ is chosen to be $ 0\!\leq\!\delta\!\leq\!\min_{k\in[0:2^{m}-1]}(b_{(k+1)}\!-\!b_k)$ for each coefficient, since at its maximum value at least half of the realizations are removed and further removal might not be practical. We observe for most transform coefficients that the decrease pattern of $\beta_j$ with respect to $P_{c,j}$ for increasing $\delta$ is different for small, medium, and large numbers $m_j$ of extracted bits. Thus, it seems difficult to obtain a general algorithm that provides optimal operation points in terms secrecy, reliability, QoS, code rate, etc. We therefore propose to extend the thresholding approaches proposed in \cite{bizimICC,bizimMDPI} by imposing thresholds on both $\beta_j$ and $P_{c,j}$, rather than only on $P_{c,j}$. The lower bound on $P_{c,j}$ is then determined by the block error probability, as defined in \cite[Eq.~(8)]{bizimICC}, and $\beta_j$ is lower bounded by a chip manufacturer as a practical constraint. Then, for the $j$-th transform coefficient for $j\in[2\!:\!r]$, the maximum number of bits that satisfies both thresholds is assigned to $m_j$ and we obtain the value $\delta_j$ that corresponds to the operation point $(P_C(\delta_j),\beta_j(\delta_j))$. One can then guarantee a QoS parameter of $\delta$ that is the minimum $\delta_j$ over all used transform coefficients, providing a guarantee for the worst case reliability of all bit sequences extracted from all used PUFs with the same hardware design.

\bibliographystyle{IEEEbib}
\bibliography{references}

\end{document}